\documentclass[aps,prd,a4paper,reprint,nofootinbib,superscriptaddress,floatfix,preprintnumbers]{revtex4-1}
\usepackage{hyperref}
\usepackage{amsmath}
\usepackage{amsfonts}
\usepackage{amssymb}
\usepackage{color}
\usepackage[utf8]{inputenc}
\usepackage{graphicx}
\usepackage{microtype}
\usepackage{siunitx}
\usepackage{subfigure}
\usepackage{soul}
\usepackage[normalem]{ulem}
\usepackage[utf8]{inputenc}
\usepackage{fontawesome}
\DeclareUnicodeCharacter{2212}{-}
\usepackage{comment}
\usepackage{wasysym}

\makeatletter
\newcommand{\github}[1]{%
   \href{#1}{\faGithubSquare}%
}
\makeatother

\newcommand{\Fermi}{\textit{Fermi}}

\begin{document}

\begin{flushleft}
LAPTH-066/22
\end{flushleft}

\title{Gamma-ray flux limits from brown dwarfs:\\ Implications for dark matter annihilating into long-lived mediators.}

\author{Pooja Bhattacharjee}\email{pooja.bhattacharjee@lapp.in2p3.fr}
\affiliation{LAPP,  CNRS, USMB, F-74940 Annecy, France}
\affiliation{LAPTh, CNRS, USMB, F-74940 Annecy, France}
\author{Francesca Calore}\email{calore@lapth.cnrs.fr}
\affiliation{LAPTh,  CNRS, USMB, F-74940 Annecy, France}
\author{Pasquale Dario Serpico}\email{serpico@lapth.cnrs.fr}
\affiliation{LAPTh, CNRS,  USMB,  F-74940 Annecy, France}

\begin{abstract}
Brown dwarfs (BDs) are celestial objects representing the link between the least massive main-sequence stars and giant gas planets. In the first part of this article, we perform a model-independent search of a gamma-ray signal from the direction of nine nearby BDs in 13 years of \Fermi-LAT data. We  find no significant excess of gamma rays, and we, therefore, set 95\% confidence level upper limits on the gamma-ray flux with a binned-likelihood approach.
In the second part of the paper, we interpret these bounds within an exotic mechanism proposed for gamma-ray production in BDs: If the dark matter (DM) of the universe is constituted of particles with non-negligible couplings to the standard model, BDs may efficiently accumulate them through scatterings. DM particles eventually thermalise, and can annihilate into light, long-lived, mediators which later decay into photons outside the BD.
Within this framework, the current Fermi-LAT stacked upper limits on the gamma-ray flux of the selected BDs do not enable setting a bound on the DM-nucleon cross-section. An improvement of approximately a factor of 9 in the gamma-ray limits is required to obtain bounds on the scattering cross-section $\sigma_{\chi n}$ $\sim 10^{-36}$ cm$^{2}$ for DM masses below 10 GeV. This improvement would cover a larger portion of the parameter space in mediator decay length and DM mass. The code and data to reproduce the results of this study are available on GitLab at \github{https://gitlab.in2p3.fr/francesca.calore/brown-dwarfs-gamma}..
\end{abstract}

\maketitle

\section{Introduction}
\label{sec:intro}
The opening of a new astronomical window is almost unavoidably accompanied by surprises. In the realm of GeV gamma rays, for instance, \Fermi-LAT found not only (largely unexpected) emission from milli-second pulsars (MSPs), but went eventually to establish MSPs as one of the dominant classes of objects of the gamma-ray sky (for a review, see~\cite{Caraveo:2013lra}). 
Even relatively familiar objects, like the Sun, have a gamma-ray emission whose spectrum and morphology still puzzles us~\cite{Nisa:2019mpb}. It is therefore important to scrutinise different classes of objects and be ready for the unexpected.

One of the {\it least} spectacular classes of astrophysical objects is constituted by brown dwarfs (BDs), 
representing the link between the least massive main-sequence stars and giant gas planets. Unlike any main-sequence star, they are not massive enough (below about 0.07-0.08 $M_\odot$) for thermonuclear fusion of hydrogen, but can burn the small quantities of deuterium that they contain. They are predicted to have similar radius as Jupiter, while weighting roughly 15 to 80 times Jupiter's mass, which is $M_{\jupiter}\simeq 1.9\times 10^{27}\,$kg, to ignite deuterium but not hydrogen fusion. 
Their existence, first invoked in the sixties~\cite{Kumar:1963a,Kumar:1963b,Hayashi:1963}, was eventually observationally confirmed three decades later~\cite{Nakajima:1995,Rebolo:1995}. Due to their low-luminosity and dominant emission in the infrared, only a few thousand of BDs have been detected, and only in our Galactic neighbourhood, even though the Milky Way may contain a billion or more of them.
Quite surprisingly, despite not being particularly flashy objects, non-thermal emission from BDs has been found, both in radio~\cite{Berger:2001rf} and X-rays~\cite{Rutledge:2000nu}. It appears to be mostly, if not exclusively, associated to flares (similarly to analogue phenomena in the comparably unimpressive red-dwarfs) and it is speculated that it might be associated to sub-surface magnetic activity. 

Although we are not aware of a ``conventional'' GeV gamma-ray production mechanism that should be expected from BDs, there is at least one exotic process that has been proposed, which could make them interesting gamma-ray targets. If the dark matter (DM) of the universe is constituted of particles $\chi$ with non-negligible couplings to the Standard Model (SM), compact astrophysical objects may efficiently accumulate them through scatterings. Typical targets considered are the Sun~\cite{Press:1985ug,Griest:1986yu,Gould:1987ir} or more compact stars, e.g.~neutron stars (NS)~\cite{Goldman:1989nd, Gould:1990b,Bertone:2007ae}, with signatures associated to heating, energy transport, black hole formation, or annihilation.  

The DM capture rate in Sun and NS has been explored in several studies \cite{Bose:2021yhz,Chen:2018ohx,Leane:2021ihh,Albert:2018jwh,Serini:2020yhb,Niblaeus:2019gjk,Leane:2017vag}. Only recently, exoplanets and BDs have been studied in this context~\cite{Leane:2020wob,Leane:2021ihh}. One comparative advantage with respect to the Sun is that BDs have a lower characteristic mass (sub-GeV) below which capture effects are counteracted by {\it evaporation effects}~\cite{Garani:2021feo}. 
Depending on the temperature and density of the celestial object, the so-called evaporation mass corresponds to the lightest DM particle which can be efficiently captured. 
Although BDs are not as efficient as NS in capturing DM with feeble interactions, for larger scattering cross sections their larger size roughly compensates their lower gravitational focus power, see eq.~\eqref{eqn:Cmax} below. Compared to NS, they are more numerous, and some very nearby ones are known: The closest BDs being at a few pc distance, to be compared with the nearest NS at hundreds of pc~\cite{Haberl:2006xe}. Overall, we thus expect that BDs may be interesting targets for light and relatively strongly-interacting DM. 

But how would one detect the impact of DM capture in BDs? One possibility, not developed further below, is to look for the anomalous heating due to the deposited kinetic energy associated to DM capture (and, possibly, to the further one due to subsequent DM annihilation); the perspectives for such searches in the James Webb Space Telescope (JWST) era have been studied in~\cite{Leane:2021ihh}. Let us just notice that an advantage compared to NS is that the BD thermal emission is already measured and much better understood, providing us with a sensible calorimetric benchmark.

Concerning indirect signals relying on remote sensing of DM annihilation products, we focus on a scenario where a new degree of freedom is present, $\phi$,  which acts as a {\it mediator} of DM interactions with the SM.  Since allowed DM-mediator couplings are typically much larger than mediator-SM ones, if $\phi$ is lighter than the DM, DM annihilation proceeds into the mediator final states. If not protected by particular symmetries, $\phi$ eventually decays into light SM particles, albeit with a lifetime that can be sizable~\cite{Martin:1997ns, HOLDOM198665, HOLDOM1986196, Okun:1982xi, Kobzarev:1966qya, Dasgupta:2020dik} and thus happen outside the star. Far from being an odd exception in the theoretical landscape, these scenarios are actually rather commonly examined in current searches for light DM particles at colliders~\cite{Abdallah:2015ter}.

The collective indirect gamma-ray signal via this mechanism from the BDs expected to populate the inner galaxy has been studied in~\cite{Leane:2021ihh}. Jupiter, the nearest giant planet, shares some of the advantages of BDs (albeit being slightly sub-optimal), and its gamma-ray signal in the same context was studied in \cite{Leane:2021tjj}, while its $e^\pm$ signal has been recently studied in~\cite{Li:2022wix}.
In the following, we bridge the gap between these two types of studies:
i) We perform (for the first time, to the best of our knowledge) a model-independent search of a gamma-ray signal from the direction of {\it known} BDs in \Fermi-LAT data. Compared to~\cite{Leane:2021ihh}, the study does not rely on extrapolations or statistical arguments on the BD population, especially towards the Galactic center (GC). ii) No excess gamma-ray signal being found, we interpret the resulting upper limits on the BDs' gamma-ray flux in a DM model annihilating into gamma rays via light, long-lived mediators. 
As we will discuss in the following, our constraints apply to a much broader range of lifetime for the mediators as compared with the result of~\cite{Leane:2021tjj}. 
We will also explicitly test that one of the main assumptions adopted in deriving previous limits, i.e. that equilibrium between capture and annihilation should be reached over the lifetime of the objects, does typically hold true for known BDs if the annihilation is not too much below current upper limits, and derive limits using the estimated age of each object. 

We organise this article as follows: In section~\ref{sec:BD_data}, we review 
the selection criteria adopted to create the BDs sample under consideration in this study. 
In section~\ref{sec:fermi_analysis}, we 
discuss how we perform the search for a gamma-ray signal from the location of our selected targets with \Fermi-LAT data, and 
present the results.
In section~\ref{sec:DMmodel}, we describe the formulation for the DM capture rate in BDs, their annihilation into long-lived mediators,
and the expected gamma-ray flux. 
We present the methodology followed to set the constraints on the DM particle parameter space in section~\ref{sec:limits_procedure}.
In section~\ref{sec:scatt_cross_section}, we derive the final limits on the DM-nucleon 
scattering cross-section. 
In section~\ref{sec:conclusions}, we compare our obtained bounds with the limits reported in recent literature studies, and we draw our conclusions.

\section{Brown dwarfs catalogue}
\label{sec:BD_data}

To date, several hundreds of BDs have been detected, mostly through the large-scale optical and near-infrared imaging surveys performed by Two-Micron All-Sky Survey (2MASS, \cite{Skrutskie:2006}) and Sloan Digital Sky Survey (SDSS, \cite{york:2000}).

We consider the BDs catalogue in  ref.~\cite{BDlist}.
In order to maximise the expected DM capture and gamma-ray flux, there are some selection requirements that it makes sense to impose:

\begin{itemize}
    \item 
We want the objects to be nearby, since the gamma-ray flux scales as the inverse
of the distance squared. The closest known BD is at about 2 pc from Earth.
For this study, we select BDs within 11 pc from us. Note that this selection criterion would be meaningful no matter which emission mechanisms were considered. 
\item The denser the BD, the larger the capture rate is expected to be.
We thus consider BDs with large masses ($M > 20 \, M_{\jupiter}$) -- the radius of BDs being about 
 $R_{\jupiter}$ for most of the objects in the catalogues. 
\item We would like to select objects which are cold and old.
A cold surface is typically associated with a cold core temperature, which is instrumental in keeping the evaporation mass low~\cite{Leane:2020wob, Leane:2021ihh, Garani:2021feo}. This means that these objects can probe the widest parameter space for light DM.
Old objects, on the other hand, are more likely to having reached equilibrium between the capture
and the annihilation rate, maximising the signal strength (see discussion in section~\ref{sec:DMmodel}). 
In general, BDs are spectrally classified in L, T, and Y types, from warmer to colder. Even though the Y-type BDs are comparatively older than the T-type ones, their masses are below 20 $M_{\jupiter}$, and we end up selecting only T-type BDs with typical estimated ages of a few Gyr.
\end{itemize}
In this way, our selected sample counts 9 T-type BDs, whose main characteristics -- of relevance for this study -- are reported in table~\ref{tab:source_datails}.
   
\begin{table*}[htbp]
\centering
\caption{Source Details: Column I: Source name and ID number for this analysis;
Column II \& III: Galactic longitude and latitude; Column IV: 
Distance to the BDs; Column V: Mass; Column VI: Radius; Column VII: Surface temperature; Column VIII: Maximum value of estimated age; Column VIII: Spectral type. Except for the age taken from different refs.~(see table), we took all the parameters from ref.~\cite{BDlist}.}
\label{tab:source_datails}
\begin{tabular}{|p{5cm}p{1.5cm}p{1.5cm}p{1.5cm}p{1.2cm}p{1.2cm}p{1.2cm}p{1.7cm}p{1.5cm}|}
\hline 
\hline
Source Name (ID number) & $\ell$ (deg) & $b$ (deg) & Distance (pc) & Mass ($M_{\jupiter}$) & Radius ($R_{\jupiter}$) & Temp. (K) & Estimated age (Gyr) & Spectral Type \\
\hline
\hline
2MASS J02431371-2453298 & 40.81 & -24.89 & 10.68 & 34 & 0.97 & 1070 & 1.7 & T6\\
(Source 1) & & & & & & & & \\
\hline
WISEPA J031325.96+780744.2 & 48.36 & 78.13 & 6.54 & 26 & 0.88 & 651 & 10 & T8.5\\
(Source 2) & & & & & & & & \\
\hline
Epsilon Indi Ba & -28.96 & -56.78 & 3.63 & 47 & 0.89 & 1276 & 3.5 & T1 \\
(Source 3) & & & & & & & & \\
\hline
SCR 1845-6357 B & -78.73 & -63.96 & 3.85 & 45 & 0.88 & 950 & 3.1 & T6 \\
(Source 4) & & & & & & & & \\
\hline
2MASS J12171110–0311131 & -175.71 & -3.19 & 10.73 & 31 & 0.95 & 870 & 10 & T7.5 \\
(Source 5) & & & & & & & & \\
\hline
WISEPC J121756.91+162640.2 A & -175.53 & 16.44 & 10.10 & 30 & 0.89 & 575 & 8 & T9 \\
(Source 6) & & & & & & & & \\
\hline
2MASS J04151954-0935066 & 63.83 & -9.59 & 5.64 & 35 & 0.91 & 750 & 10 & T8 \\
(Source 7) & & & & & & & & \\
\hline
2MASS J09373487+2931409 & 144.39 & 29.53 & 6.12 & 58 & 0.79 & 810 & 10 & T7 \\
(Source 8) & & & & & & & & \\
\hline
WISE J104915.57-531906.1 & 162.33 & -53.32 & 2 & 33.5 & 0.85 & 1350 & 4.5 & T0.5 \\
(Source 9) & & & & & & & & \\
\hline
\hline
\end{tabular}
\end{table*}

\section{Search for gamma-ray emission in \Fermi-LAT data}
\label{sec:fermi_analysis}
\subsection{\Fermi-LAT data selection and model}
The \Fermi-LAT is a gamma-ray space-based detector that scans the whole sky every 3 hours for an
collecting photons from about 20 MeV to almost 1 TeV. 
We here analyse nearly 13 years (2008-08-04 to 2021-12-14) of \Fermi-LAT data from the direction of the
9 T-type BDs selected in section~\ref{sec:BD_data}.

For our analysis, we use \textit{fermipy}, version 1.0.1, and \textit{Fermi Science Tools}, version 2.0.8,
that allow us to avail the latest source class instrument response function (IRF), i.e.~$\rm{P8R3\_SOURCE\_V2}$ processed data. 
We consider photons in the energy range  $E\in [0.1, 500]$~GeV and within a $15^{\circ}$ region of interest (ROI) around the position of each BD -- following the main recommendations for point-like source analysis provided by the Fermi-LAT Collaboration.\footnote{\url{https://fermi.gsfc.nasa.gov/ssc/}}
We binned the data in 24 energy bins and spatial pixels of 0.1$^{\circ}$ side. Data are prepared following LAT recommendations on data quality and selection, see table~\ref{tab:fermi_parameters}.

In each ROI, independently, we perform a binned likelihood analysis to fit the observed counts 
to the gamma-ray 
sky model of the very same ROI.
The model is composed of standard gamma-ray fore- and back-grounds: 
The Galactic ($\rm{gll\_iem\_v07.fits}$) and 
isotropic ($\rm{iso\_P8R3\_SOURCE\_V2\_v1.txt}$) diffuse emissions with free normalisation, the emission from all point-like sources present in an extended $20^{\circ}$ ROI and catalogued in the 4FGL-DR3 catalogue.  
Including sources beyond the fitted ROI is recommended given the poor angular resolution 
at low energies, which can cause some leakage of photons into the ROI from sources just outside it.
On the other hand, in order to guarantee the stability of the fit and limit the number of free parameters,
the spectral parameters of all 
4FGL point-like sources within $10^{\circ}$ from the ROI center are free to vary, while we fix them to the 4FGL catalogue values for sources beyond $10^{\circ}$.

In addition, we model the gamma-ray emission from each BD as a new point-like source at the BD position.
For this, the signal spectrum in each energy bin is modelled by a 
power-law (PL) spectrum (i.e. ${\rm d}N/{\rm d}E \propto E^{-\Gamma}$) with spectral index 
$\Gamma$ = 2.

The model counts are obtained by convolving the model components with the IRF corresponding 
to the data-set used, which accounts for energy and angular resolution.

\subsection{Results and flux upper limits}
We quantify the evidence for an excess emission from the BD point-like source,
on top of the fore-/back-ground model, through a Test Statistics defined as TS$= -2\ln(L_{\rm max,0}/L_{\rm max, 1})$, where $L_{\rm max, 0}$ and $L_{\rm max, 1}$ represent the maximum likelihood for the null hypothesis (fore-/back-ground only) and in the presence of the additional source, respectively.

No significant excess emission is found at the location of any BD.
For all BDs, TS values are well below the point source threshold detection limit TS$=25$. The maximum local (i.e.~without accounting for trial factor) significance among our sources is 2.2$\sigma$.

In the absence of an excess, for each BD, we derive the
gamma-ray flux upper limits at 95$\%$ confidence level (C.L.) using the profile likelihood method~\cite{Rolke:2004mj}. In this method, the fitting will be continued until the difference of the log-likelihood, i.e.~$-2\Delta\ln(\mathcal{L})$ function reaches the value 2.71 which corresponds to the one-sided 95$\%$ C.L..

\begin{figure*}[htbp]
 { \includegraphics[width=1.0\linewidth]{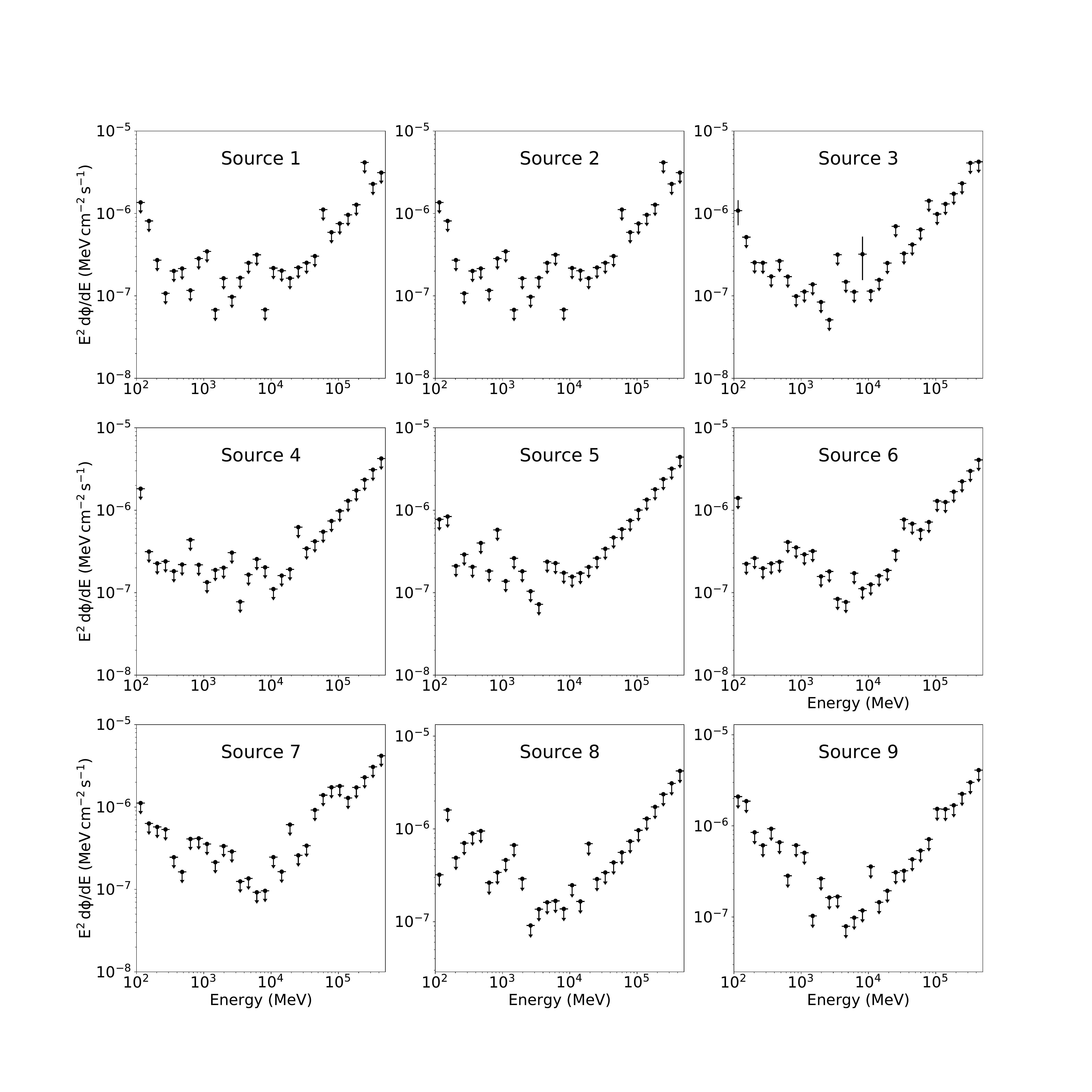}}
 \caption{Bin-by-bin $E^2 {\rm d}\Phi/{\rm d} E$ upper limits at 95\% C.L. for our selected BDs. 
}
 \label{fig:diff_flux_limits}
\end{figure*}

In figure~\ref{fig:diff_flux_limits}, we show the observed bin-by-bin $E^2~{\rm d}\Phi/{\rm d}E$ flux upper limits for all 9 BDs considered in this work.

We checked that the flux upper limits are robust with respect to background modelling uncertainties by repeating the analysis for a $5^{\circ}$ ROI, finding negligible differences.
Also, we performed the analysis with two alternative PL index values, e.g.~$\Gamma=2.5$ and $\Gamma=0$ \footnote{$\Gamma=0$ is motivated as a proxy of the DM box spectrum as discussed below, whereas we use $\Gamma=2.5$ to check the constraints on a number astrophysical source spectra, which are steeper to the limiting \Fermi~spectrum $\Gamma=2$.}, and we did not find any significant difference, as expected by the energy-binned likelihood approach.

\begin{table*}
    \caption{Parameters used for the analysis of \Fermi-LAT data.}
    \label{tab:fermi_parameters}
    \begin{tabular}{||p{7 cm}p{8 cm}||}
        \hline \hline
        Parameter & Value\\
        \hline \hline
        Radius of interest (ROI) &  $15^{\circ}$\\
        TSTART (MET) & 239557418 (2008-08-04 15:43:37.000 UTC)\\
        TSTOP (MET) & 661205870 (2021-12-14 20:17:45.000 UTC)\\
        Energy Range & 100 MeV - 500 GeV\\
        \textit{Fermitools} version & \texttt{2.0.8}\\ 
        \textit{fermipy} version & \texttt{1.0.1}\\ 
        \hline \hline
        \texttt{gtselect} for event selection &\\ 
                \hline \hline
        Event class & Source type (128)\\ 
                Event type & Front+Back (3)\\ 
        Maximum zenith angle cut & $90^{\circ}$\\
        \hline \hline
        \texttt{gtmktime} for time selection &\\ 
                \hline \hline
        Filter applied & $\textit{(DATA\_QUAL$>$0)\&\&(LAT\_CONFIG==1)}$\\ 
                ROI-based zenith angle cut & No\\ 
        \hline \hline
        \texttt{gtltcube} for livetime cube &\\ 
                \hline \hline
        Maximum zenith angle cut ($z_{cut}$) & $90^{\circ}$\\ 
                Step size in $cos(\theta)$ & 0.025\\
        Pixel size (degrees) & 1\\
        \hline \hline
        \texttt{gtbin} for 3-D counts map &\\ 
                \hline \hline
        Size of the X $\&$ Y axis (pixels) & 140\\
        Image scale (degrees/pixel) & 0.1\\
        Coordinate system & Celestial (CEL)\\
        Projection method & AIT\\
        Number of logarithmically uniform energy bins & 24\\ 
        \hline \hline
        \texttt{gtexpcube2} for exposure map &\\ 
                \hline \hline
        Instrument Response Function (IRF) & $\rm{P8R3\_SOURCE\_V2}$\\ 
                Size of the X and Y axis (pixels) & 400\\
        Image scale (degrees/pixel) & 0.1 \\
        Coordinate system & Celestial (CEL)\\
        Projection method & AIT\\
        Number of logarithmically uniform energy bins & 24\\ 
        \hline \hline
        Diffuse and source model XML file &\\ 
                \hline \hline
        Galactic diffuse emission model & $\rm{gll\_iem\_v07.fits}$\\ 
                Extragalactic isotropic diffuse emission model & $\rm{iso\_P8R3\_SOURCE\_V2\_v1.txt}$\\
        Source catalog & 4FGL-DR3 \\
        Spectral model &  Power Law with index, $\Gamma$= 2\\ 
        \hline \hline
    \end{tabular}
\end{table*}

\section{DM capture in BDs and gamma-ray emission}
\label{sec:DMmodel}
In this section, we highlight the main ingredients that enter the determination 
of the gamma-ray flux expected from DM captured in BDs.
As mentioned in section~\ref{sec:intro}, we focus on the case in which DM scatters off  nucleons (possibly, electrons),
becoming gravitationally bound to the BD and accumulating therein.
This occurs if the DM mass is not too light so that  
the velocity attained at thermalisation stays safely below the BD escape velocity, in order to prevent evaporation.  We will consider the scattering cross section and the mediator lifetime as independent phenomenological parameters. The rationale for that is twofold: i) The DM-$\phi$ coupling, entering the scattering and annihilation but not the $\phi$ decay, is only poorly constrained and potentially large; while the SM-$\phi$ (entering the DM scattering as well as the $\phi$ lifetime) is only subject to upper limits. ii) Our choice also encompasses the case where the DM capture and thermalisation is mediated by some effective operator depending on high-scale physics (well above DM mass), and the coupling with the mediator $\phi$ affects only the annihilation process. The medium being constituted in large measure by protons (as opposed to nuclei), we are almost equally sensitive to spin-dependent as to spin-independent cross section, a situation very different from the direct DM searches in the laboratory. 

The calculation can be divided into two major steps:
i) The capture of DM into BDs, which depends notably on the DM local density and velocity distribution, as well as scattering cross section of DM. 
ii) The annihilation signal of the DM gravitationally trapped inside BDs, and the ensuing gamma-ray emission from the decay of long-lived mediators.

\subsection{Dark matter capture in BDs}
When DM particles transit through the celestial object, if colliding one or multiple times,  they can lose sufficient energy and eventually get captured. The \textit{``maximum capture rate''} ($C_{\rm max}$) is defined by the condition that all the DM particles passing through the cross section of the body get captured. A naive estimate for the maximum capture rate of a body of radius $R_\star$ in a DM environment with number density $n_{\chi}=\rho_\chi/m_\chi$, and with relative velocity, $v_0$ would be~\cite{Leane:2020wob}:
\begin{equation}
C_{\rm max}\simeq \pi R_{\star}^2 n_\chi v_0\,.\label{eq:simpleCmax}
\end{equation}
If we denote with $\bar{v}$ the root mean square of the DM velocity distribution (also known as velocity dispersion), then  $v_{0} = \sqrt{8/(3\pi)}\bar{v}$. The quantity $\bar{v}$ is related to the circular velocity at any galactocentric distance $r$, by the relationship $\bar{v}(r) = 3/2 v_{\rm c}(r)$, where $v_{\rm c}(r) = \sqrt{G M(r)/r}$, $G$ being the gravitational constant and $M(r)$ the total Galactic mass enclosed within $r$. Note that, although in the following we concentrate on BDs in the solar system neighbourhood, for validation purposes we implement a model valid for any point in the Galaxy. For the mass model of the Milky Way, we follow the ref.~\cite{Sofue:2013}, and we include the mass of the supermassive black hole, the exponential disk, the inner spheroidal bulge, the outer spheroidal bulge, and DM halo density.
All the BDs in our sample being less than 11 pc from Earth, we fix $r \equiv R_{\odot} =$ 8.1 kpc \cite{2021ARep...65..498B}, so that $v_0 = 293.4$ km/s.
For the DM density distribution, we assume, if not stated otherwise,
a typical Navarro-Frenk-White profile with parameters as in~\cite{Calore:2022stf}, which gives us 
$\rho_\chi(r = R_\odot) \equiv \rho_0 = 0.39$ GeV/cm$^{3}$ (i.e.~0.0103 $M_\odot$/pc$^3$).

The eq.~\eqref{eq:simpleCmax} does not account for gravitational focusing of the celestial body. Its generalisation including that is~\cite{Garani:2017jcj}: 
\begin{equation}
C_{\rm max}(r) = \pi R_\star^{2} n_{\chi}(r) v_{0}\, \left( 1+ \frac{3}{2} \frac{v^{2}_{\rm esc}}{\bar{v}(r)^{2}} \right) \, ,
\label{eqn:Cmax}
\end{equation}
where 
$v_{\rm esc}$ = $\sqrt{2GM_{\star}/R_{\star}}$ is the escape velocity (with $M_{\star}$ the mass of the celestial body, and $R_{\star}$ its radius).
Although the correction term in eq.~\eqref{eqn:Cmax} is small for the velocities at play here, we nonetheless
use this equation to derive the maximal capture rate in what follows.

The eq.~\eqref{eqn:Cmax} only applies if all the  DM particles passing through are captured by the BDs. If, as often happens, this is far from being the case, a perturbative approach can be followed. In this paper, we follow the formulation of ref.~\cite{Bramante:2017xlb, Ilie:2020}, accounting for both single and multiple scatterings of DM particles. The probability of DM scattering $N$ times before getting captured is:
\begin{equation}
p_{N}(\tau)=2\,\int_{0}^{1}\, dy \,\frac{y\,e^{-y\tau}\,(y\tau)^N}{N!}
\label{eqn:p_N}
\end{equation}
Here $\tau$ is the optical depth, namely $\tau = \frac{3}{2} \frac{\sigma_{\chi n}}{\sigma_{\rm sat}}$, where $\sigma_{\chi n}$ is the DM-nucleon scattering cross section; $\sigma_{\rm sat}$ = $\pi R_\star^{2}/N_{n}$ the saturation cross section; $N_{n}=M_\star/m_n$ is the number of nucleons in the target, with $m_{n}$ the nucleon mass. 
The expression of the total capture rate ($C_{\rm tot}$) is now:

\begin{equation}
 C_{\rm tot}(r) = \sum_{N=1}^{\infty} C_{N}(r)
 \label{eqn:c_tot}
\end{equation}
where the capture rate associated with  $N$ scatterings, $C_{N}$, writes
\begin{widetext}
\begin{eqnarray}
C_{N}(r) &=& \pi\,R_\star^{2}\,p_{N}(\tau)\, \frac{\sqrt{6}n_{\chi}(r)}{3\sqrt{\pi}\bar{v}(r)}\, \times \,\left[(2\bar{v}(r)^{2}\,+\,3v_{\rm esc}^{2})\,-\,(2\bar{v}(r)^{2}\,+\,3v_{N}^{2})\,\exp\left(-\frac{3(v_{N}^{2}\,-\,v_{\rm esc}^{2})}{2\bar{v}(r)^{2}}\right)\right]\,.
\label{eqn:c_N}
\end{eqnarray}
\end{widetext}

In eq.~\ref{eqn:c_N}, $v_{N}$ is defined as $v_{N}~=~v_{\rm esc}(1-\beta_{+}/2)^{-N/2}$ where $\beta_{+}~=~4m_\chi m_{n}/(m_\chi+m_{n})^{2}$. In case of large number of scatterings ($N\gg$1), ($v_{N}^{2}\,-\,v_{\rm esc}^{2}$) dominates over $\bar{v}^{2}$ with $C_{N}$ reducing to $\approx p_{N} C_{\rm max}$. In relation to $C_{\rm tot}$ and $C_{\rm max}$, we check that the perturbative estimate is applicable here, so we impose the condition $C={\rm min}[ C_{\rm tot}, C_{\rm max}]$. 

We remind the reader that, for low DM masses, the evaporation of DM particle becomes important. 
In ref.~\cite{Leane:2020wob}, the evaporation mass in BD was estimated to be around a few MeV. However, a recent study~\cite{Garani:2021feo} estimated an evaporation mass $\sim$0.7 GeV. 
In our paper, we do not recompute the evaporation mass, nor include it in the number density evolution equation. Yet, consistently with~\cite{Garani:2021feo}, we limit ourselves to masses above 0.7 GeV, where this approximation should be good.

\subsection{Dark matter annihilation in BDs}
\label{sec:dm_annihilation}
After getting trapped in a BD, if self-annihilations are allowed, these will contribute to determine the maximum number of DM particles. Indeed, neglecting evaporation, the number of DM particles $N(t)$ accumulated in the core of the celestial object at time $t$ obeys the equation:
\begin{equation}
\frac{{\rm d}N(t)}{{\rm d}t} = C - C_{\rm ann}N^2(t)\,, \\
\label{eqn:N_variation}
\end{equation}
where $C$ is the total capture rate previously discussed, and $C_{\rm ann}$ is the annihilation rate, further detailed below.
In the regime where evaporation is not efficient, the solution to eq.~\ref{eqn:N_variation} writes as: 
\begin{equation}
N(t) = C t_{\rm eq} \tanh\frac{t}{t_{\rm eq}} \label{eqn:Neq_time}
\end{equation}
where
\begin{equation}
t_{\rm eq} \equiv \frac{1}{\sqrt{C_{\rm ann}C}}\,, 
\label{eqn:equilibrium_time}
\end{equation}
is the equilibration time in absence of evaporation. 
The annihilation rate is defined as:
\begin{equation}
C_{\rm ann} = \langle\sigma_{\rm ann}v\rangle/V_{\rm ann} \, ,
\label{eqn:Cann}
\end{equation}
where $\langle\sigma_{\rm ann}v\rangle$ is the thermally averaged annihilation cross section,
and $V_{\rm ann}$ is the containment volume where annihilation takes place, mostly confined in the inner region of the BD.
Following~\cite{Li:2022wix} and refs. therein, we write the annihilation volume as: 
\begin{eqnarray}
V_{\rm ann} &=& \frac{4}{3} \pi r_{\rm ann}^3 = \frac{9}{\sqrt{4 \pi}} \left( \frac{T_{\star, c}}{ G \rho_{\star, c} m_\chi}\right)^{3/2}  \nonumber\\
&\simeq& 0.0087 R_{\jupiter}^3 \left( \frac{T_{\star, c}}{2 \times 10^{5} \, \rm K} \frac{200 \, \rm g/cm^3}{\rho_{\star, c}}\frac{1 \, \rm GeV}{m_\chi} \right)^{3/2} 
\label{eqn:Vann_scale} 
\end{eqnarray}
with $T_{\star, c}$ and $\rho_{\star, c}$ being the temperature and density in the body interior, see e.g.~\cite{Bramante:2017xlb}, within which the DM particle are assumed to be homogeneously distributed. 
We obtain $T_{\star, c}$ and $\rho_{\star, c}$ of our selected BDs by following the analytical formula in ref.~\cite{Auddy:2016abc},
and  we report the values in table~\ref{table:bd_Trho}.

\begin{table}
\centering
\caption{Values of $T_{\star, c}$ and $\rho_{\star, c}$ for the 9 BDs studied here.\label{table:bd_Trho}}
\label{Tab-3}\begin{tabular}{|p{2cm}p{2cm}p{2cm}|}
\hline 
\hline
Source & $T_{\star, c}$~[$\times 10^5$ K] & $\rho_{\star, c}$~[g/cm$^{3}$] \\
\hline
\hline
1 & 3.35  & 248.94 \\
\hline
2 & 1.40  & 145.57 \\
\hline
3 & 4.19  & 475.70 \\
\hline
4 & 4.53  & 436.10 \\
\hline
5 & 2.22 & 206.95 \\
\hline
6 & 1.87  & 193.82 \\
\hline
 7 & 2.27  & 263.80 \\
\hline
 8 & 5.11 & 724.42 \\
\hline
9 & 2.58  & 241.67 \\
\hline
\hline
\end{tabular}
\end{table}

In general, the total DM annihilation rate, $\Gamma_{\rm ann}$, writes:

\begin{equation}
\Gamma_{\rm ann}(t) \equiv \frac{C_{\rm ann}N^2(t)}{2} = \frac{C}{2} \left(\tanh \frac{t}{t_{\rm eq}} \right)^2
\label{eqn:cap_ann_age}
\end{equation}
and it is a function of, among other parameters, the DM particle mass and scattering cross section. 
We notice that the factor of 2 comes from the fact that two DM self-conjugate particles participate in each annihilation process -- this factor is 4 for non-self-conjugate DM particles.

If equilibrium is reached today, i.e.~$t_\star \gg t_{\rm eq}$, the total annihilation rate only depends on the
capture rate: 
\begin{equation}
\Gamma_{\rm ann} \rightarrow \frac{C}{2}\, .
\label{eqn:cap_ann}
\end{equation}

Note that, unlike the diffuse DM signal from annihilations in the halo, if equilibrium holds the DM annihilation rate is proportional to the DM density, i.e. $\Gamma_{\rm ann}~\propto~n_{\chi}$ (as opposed to $\Gamma_{\rm ann}~\propto~n^{2}_{\chi}$). For a population of compact bodies with density $n_\star$, $\Gamma_{\rm ann}~\propto~n_{\chi}n_{\star}$.

We will discuss below the extent to which the 
equilibrium assumption is justified, given the 
estimated age of the {\it known} BDs selected.

\subsection{Dark matter gamma-ray spectrum}
\label{sec:dm_spectrum}

For simplicity and in order to ease comparison with the literature, we will consider the single channel $\phi\to\gamma\gamma$, although it is straightforward to generalise to gamma rays emitted via other final states. 
Note that the bounds only apply when the decay range $L$ is much longer than $R_\star$, and an upper limit to the range also follows from technical hypothesis on the data-analysis, as detailed in the following paragraph. We also work under the approximation that $m_\phi\ll m_\chi$: Although the formulae reported below are valid also at a finite value of $m_\phi$, this approximation frees ourselves from the dependence from the extra parameter $m_\phi$. In a more extended scan of the parameter space, it could be easily lifted, of course. Note that the decay length is related to the decay width $\Gamma$ of the mediator $\phi$ via $L= m_\chi/(m_\phi\Gamma)$.

For practical limitations, we also want to limit ourselves to consider the gamma-ray signal emitted from BDs as point-like. Otherwise, extended source analysis will have to be performed, and eventually, the signal may become too broad on the sky to derive any meaningful constraints in the typical ROI. 
This condition means that $\phi$ must decay within a distance from the BD such that, seen from the typical few pc distance from us, its subtended angle is smaller than the angular resolution of the LAT. The minimum 68\% containment radius in the energy range considered at the BD position is assumed to be $\theta_{68\%} = 0.1^\circ$. These conditions restrict 
\begin{equation}
10^8\,{\rm m}\simeq R_\star \lesssim L \lesssim d_\star\,\theta_{68\%}\simeq 10^{14}\,{\rm m}\,.\label{decay_range}
\end{equation}
Note that analogous conditions applied to Jupiter, as in~\cite{Leane:2021tjj}, lead to a similar lower limit but to a five orders of magnitude more stringent upper limit, controlled by the significant hierarchy in distance between $\sim 5$AU (typical distance to Jupiter) and a few pc (distance to the nearest BDs). 
Hence, even when our bounds are not naively competitive to those reported in~\cite{Leane:2021tjj}, they apply to much wider parameter space. 

Within the above assumptions, the expression of the DM gamma-ray flux at Earth expected from long-lived mediators decay is
\begin{eqnarray}
E^{2}\frac{{\rm d}\Phi}{{\rm d}E} &=& \frac{\Gamma_{\rm ann}}{4 \pi d_\star^{2}}~\times~E^{2}\frac{{\rm d}N}{{\rm d}E} \,.
\label{eqn:dm_flux}
\end{eqnarray}

The gamma-ray flux coming from the location of BDs will behave as the point-like gamma-ray source and under that assumption, the gamma-ray spectra (resulting from $\phi~\rightarrow~\gamma \gamma$) will form a box-shaped spectrum:
\begin{equation}
\frac{{\rm d}N}{{\rm d}E} = \frac{4\Theta(E-E_{-}) \Theta(E_{+}-E)}{\Delta E}\,,
\label{eqn:dm_spectra}
\end{equation}
where $\Theta$ is the Heaviside function, $E_{\pm}\equiv(m_\chi/2)(1\pm \sqrt{1-m^{2}_{\phi}/m_\chi^2})$ are the box edges, and $\Delta E\equiv E_{+}-E_{-}=\sqrt{m_\chi^2-m^{2}_{\phi}}$ is the box width. Under the assumption $m_{\phi}\ll m_\chi$ the dependence on $m_{\phi}$ is lost and the expression of the photon spectrum simplifies into:
\begin{equation}
\frac{{\rm d}N}{{\rm d}E} \simeq \frac{2\Theta(m_\chi-E)}{m_\chi} \,.
\label{eqn:dm_spectra_lim}
\end{equation}

\section{Scattering cross-section upper limits}
\label{sec:limits_procedure}
From the gamma-ray flux upper limits derived in section~\ref{sec:fermi_analysis}, we can derive constraints on 
the DM parameter space as follows.
For a chosen value of the DM mass, we determine the DM spectrum from eq.~\eqref{eqn:dm_spectra_lim}.
By comparing then eq.~\eqref{eqn:dm_flux} with the bin-by-bin upper limits in figure~\ref{fig:diff_flux_limits}, 
we can derive the upper limit value on $\Gamma_{\rm ann}$, and ultimately on the scattering cross section.

For the binned LAT likelihood analysis, we define the likelihood function as the product of the probabilities of 
observing the predicted counts in each bin.
For the $i^{\rm th}$ BD, this reads:
\begin{equation}
\mathcal{L}_{i} =  \prod \limits_{j} \frac{\lambda_{j}^{n_{j}} e^{-\lambda_{j}}}{n_{j}!}= e^{-N_{\rm pred}} \prod \limits_{j} \frac{\lambda_{j}^{n_{j}}}{n_{j}!}
 \label{eqn:L_single}
\end{equation}
where, $N_{\rm pred}$ denotes the total number of predicted counts from the source model. The procedure applied to each BD in the sample yields individual BD upper limits.

One can also combine the null detections from all BDs into a stacked analysis via a conventional joint likelihood method.
The joint likelihood function is the product of the 
individual likelihoods 
 i.e.~$
\mathcal{L}_{\rm stack} =\prod \limits_{i} \mathcal{L}_{i}$.
Each BD flux contribution shares the same DM parameters, while distance, mass, and radius of each source, as well as the background within each ROI, are source-specific. 
Since here we do not have nuisance parameters to profile over, the combination is fairly straightforward. 

\section{Results: DM cross section upper limits}
 \label{sec:scatt_cross_section}

In this section, we translate the \Fermi-LAT gamma-ray flux upper limits (cf.~figure~\ref{fig:diff_flux_limits}) into constraints on the DM particle parameter space. 
In particular, we set bounds on the scattering cross section of DM particles with nucleons, $\sigma_{\chi n}$, 
as a function of the DM mass, $m_\chi$.
We compute upper limits for each BD in our selection, as well as from the combined sample (stacking procedure).
We do that, first under the hypothesis that equilibrium has been reached within the lifetime of the BD, 
and then by properly accounting for the real age of each of the objects.

\subsection{Equilibrium hypothesis}
\begin{figure}
\centering
\includegraphics[width=0.85\linewidth]{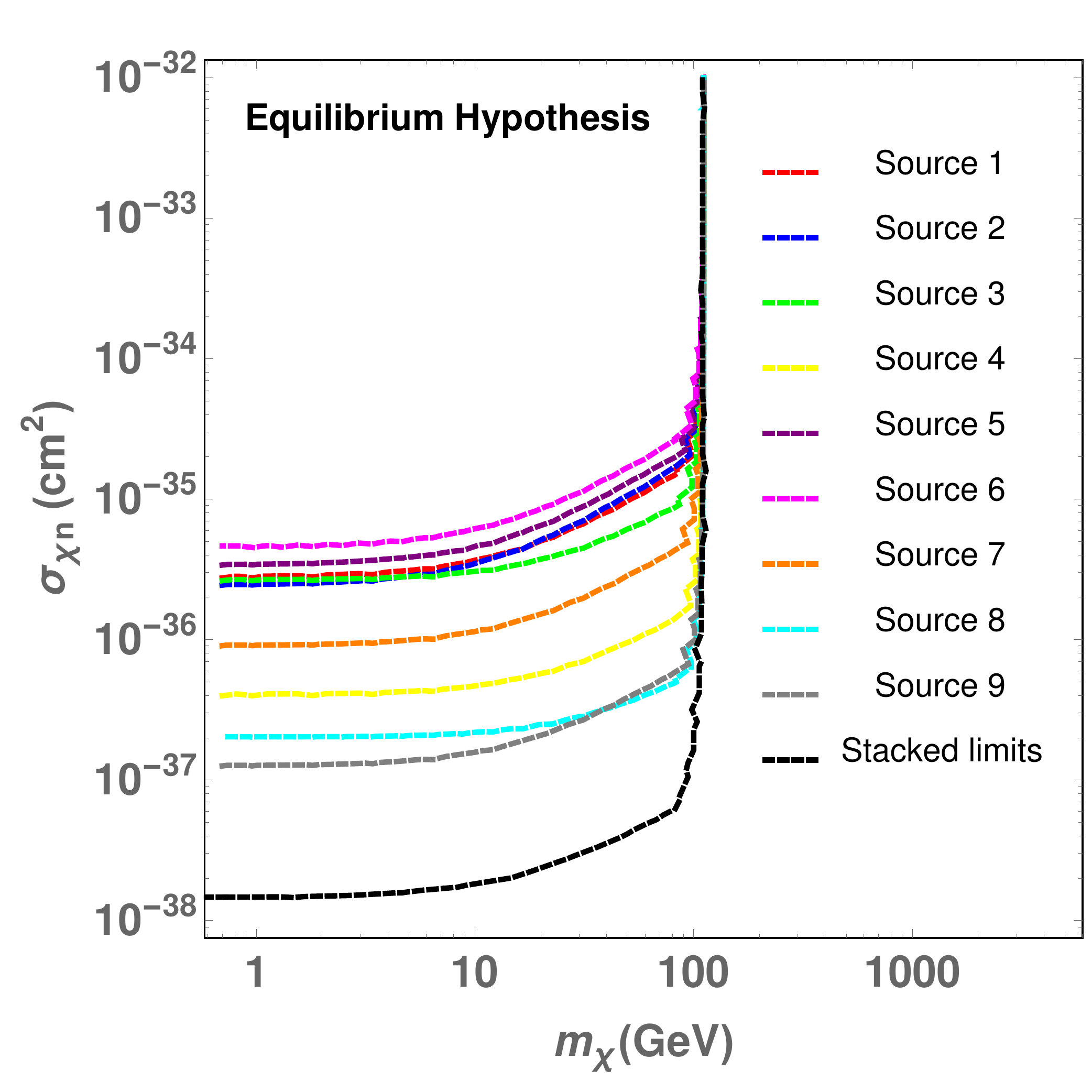}
\caption{Individual BD and stacked scattering cross section upper limits as a function of DM mass, under the hypothesis that equilibrium has been reached within the BD lifetime.
}
\label{fig:stack}
\end{figure}

Figure~\ref{fig:stack} shows the individual scattering cross section limits for the mediators decaying into gamma rays, via $\chi\chi~\rightarrow~\phi \phi\to 4\gamma$ for all 9 BDs separately, together with the stacked limit. 

In general, the limits depend on the BD characteristic properties (distance, mass and radius) in a rather intuitive way: BDs with comparable $R_\star$ and $d_\star$ yield bounds that are more stringent than the larger $M_\star$ is, (i.e. the more compact the BD is). Similarly, at fixed $d_\star$ and $M_\star$, the bounds are more stringent the smaller the radius $R_\star$, again corresponding to a more compact BD. Finally, at fixed $R_\star$  and $M_\star$, the flux scales simply as $d_\star^{-2}$ and the constraints follow correspondingly,  modulo statistical fluctuations and different background intensity in different directions.

Also, note that the stacked limit improves by at least a factor of $\sim 9$ with respect to the limits obtained from source 9 (the best among all the 9 sources). We checked that removing a few of the least luminous ones does not appreciably alter the stacked bound.

We would like to stress here that 
the dependence of the bounds on the adopted profile of DM is modest.
Its main dependence is the proportionality to the local density of DM, $\rho_0$, which amounts to $\approx$ 30\% if one uses for example a Burkert profile with parameters as in~\cite{Calore:2022stf}.
The error on $\rho_0$ is the largest theoretical uncertainty at play here. This is however common to all local probes of DM, including direct detection experiments.

\subsection{Real age}
\label{sec:limits_age}
We discuss now what is the impact of considering the real age when computing 
the DM annihilation rate, following eq.~\eqref{eqn:cap_ann_age}.
In this case, we fix the annihilation cross section to the 
current upper limits from Planck~\cite{Leane:2018kjk}, which are the strongest in the considered
DM mass range. In particular, we use the result   $\langle \sigma v \rangle \lesssim 5.1 \times 10^{-27} (m_\chi/\rm GeV)\,  \rm cm^3/s$, valid for $m_\chi \lesssim 5\,$GeV.

In general, the equilibrium time for a 1 GeV mass DM particle can be written as: 
\begin{eqnarray}
t_{\rm eq} &= 1.24 \, t_{\rm eq}^{\jupiter} \times \left( \frac{T_{\star, c}}{2 \times 10^{5} \, \rm K} \frac{200 \, \rm g/cm^3}{\rho_{\star, c}} \right)^{3/4} \times \nonumber \\
& \sqrt{\frac{5 \times 10^{-27} \rm cm^3/s}{\langle \sigma v \rangle}} \sqrt{\frac{10^{-38} \rm cm^2}{ \sigma_{\chi n} }} \, .
\label{eqn:equilibrium_time_scale}
\end{eqnarray}

We estimate typical BD's equilibrium times ($t_{\rm eq}$) of $\mathcal{O}(100)$ Myr, to be compared with 0.3 Gyr for Jupiter, according to~\cite{Li:2022wix}.
In table~\ref{table:bd_equlibrium}, we report  $t_{\rm eq}$ for $m_{\chi}$=1 GeV and $\sigma_{\chi n}$=$10^{-38}$~$\rm cm^{2}$.
These values of $t_{\rm eq}$  are 
 comparable or shorter than the estimated ages of our BDs reported in table~\ref{tab:source_datails}, especially for the coolest stars. Hence, the equilibrium hypothesis is well justified for the objects under study, provided that the annihilation cross sections are not much smaller than the upper limit considered.

\begin{table}
\centering
\caption{Equilibrium time scale ($t_{\rm eq}$) for BDs at $m_{\chi}$ = 1 GeV and $\sigma_{\chi n}=10^{-38}$~$\rm cm^{2}$.\label{table:bd_equlibrium}} 
\label{Tab-3}\begin{tabular}{p{2cm}p{2cm}}
\hline 
\hline
Source & $t_{\rm eq}$~[Gyr] \\
\hline
\hline
 1 &  0.46 \\
 2 &  0.36 \\
 3 & 0.33 \\
 4 &  0.38 \\
 5 &  0.39 \\
 6 &  0.36 \\
 7 &  0.33 \\
 8 &  0.28 \\
 9 &  0.39 \\
\hline
\hline
\end{tabular}
\end{table}

\begin{figure}
\centering
\includegraphics[width=0.85\linewidth]{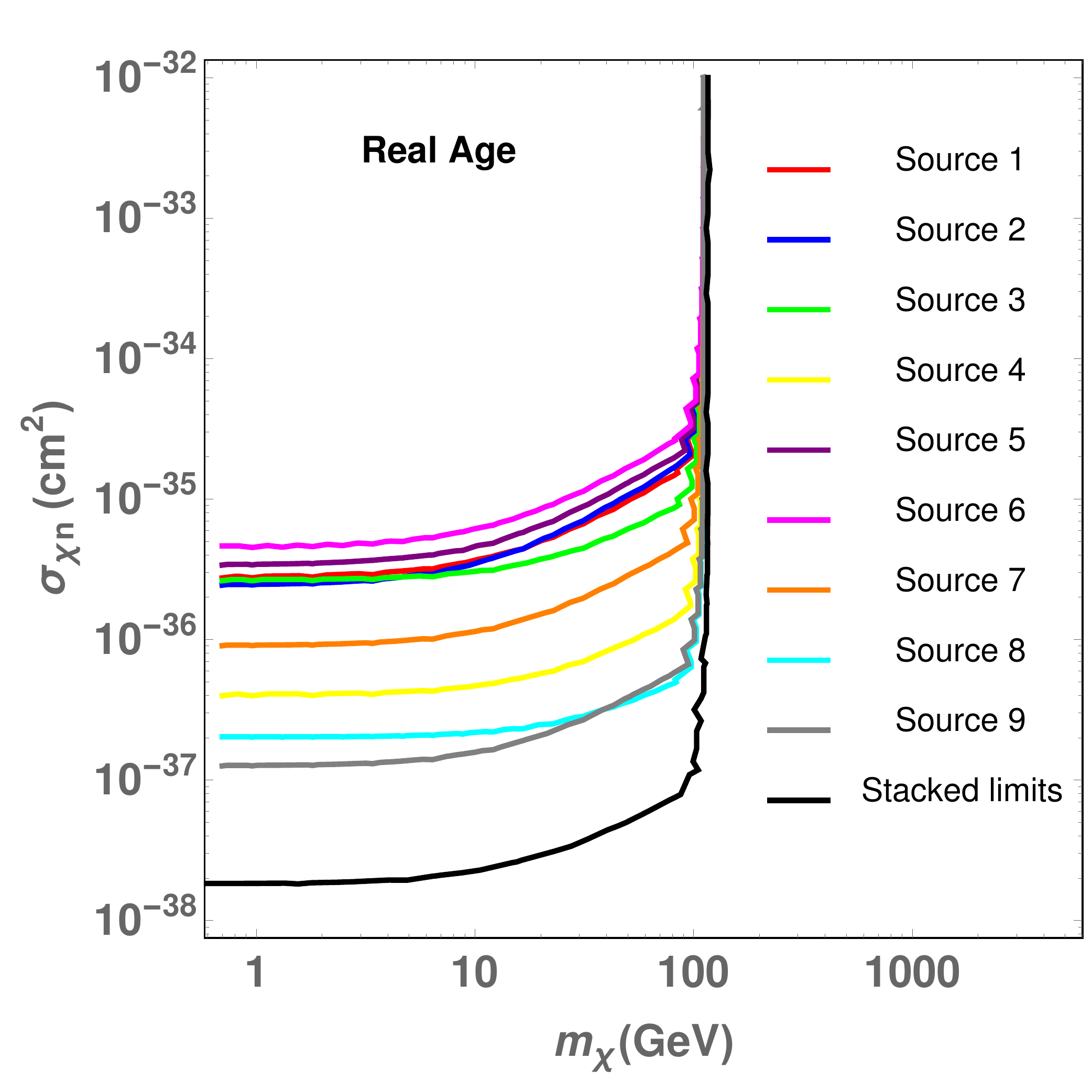}
\caption{Individual BD and stacked scattering cross section upper limits as a function of DM mass, using the real age in table~\ref{tab:source_datails} for each BD in the sample.
} 
\label{fig:limits_age_stack}
\end{figure}

\begin{figure}
\centering
\includegraphics[width=0.85\linewidth]{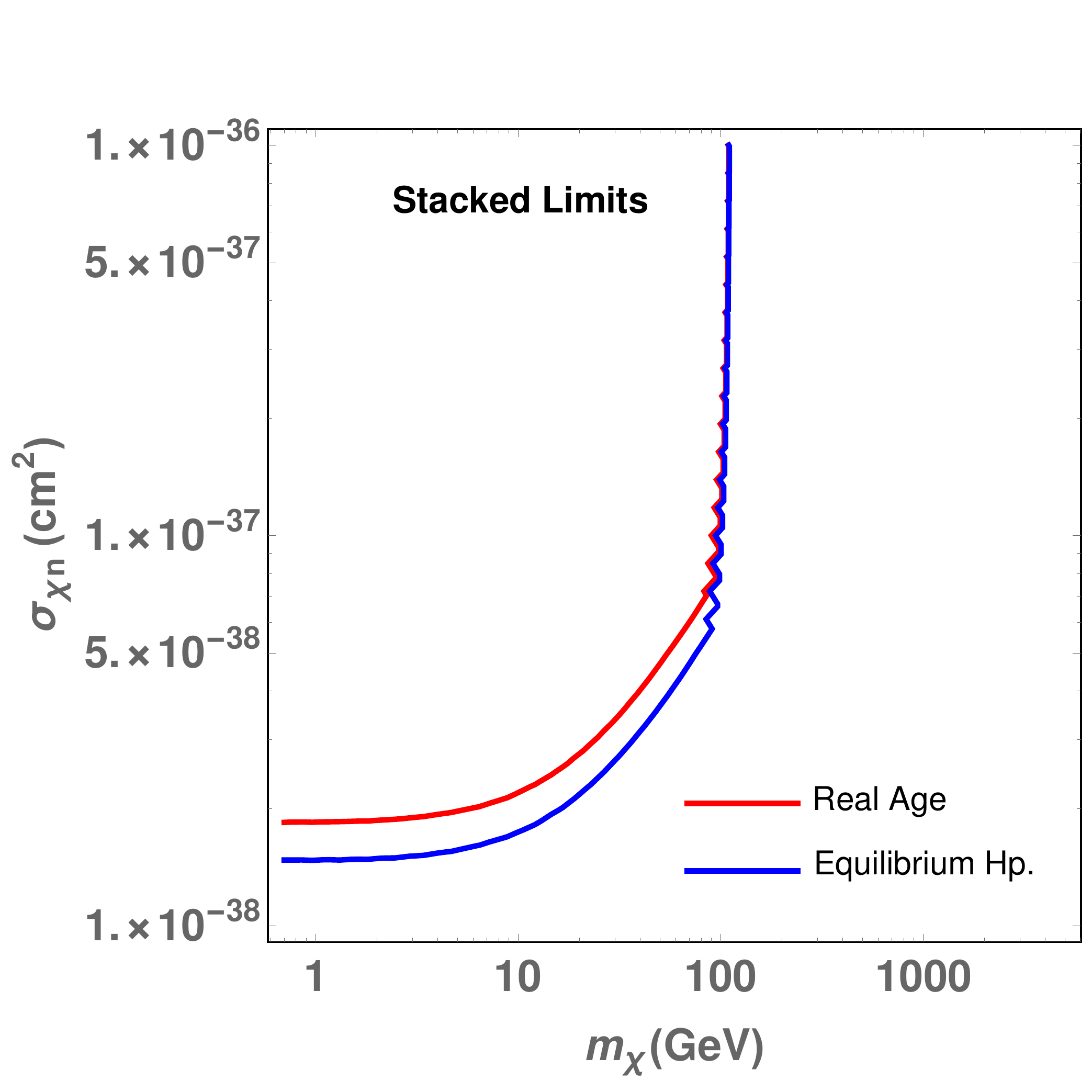}
\caption{The comparison between the stacked scattering cross section limits obtained from 1) assuming the equilibrium and 2) real age.}
  \label{fig:limits_age_comparison}
\end{figure}

We display the limits derived for the single BDs and the combined case in figure~\ref{fig:limits_age_stack}.
As expected from the $t_{\rm eq}$, the bounds obtained from considering the real BDs age are only mildly weaker from the ones where equilibrium is assumed, but are nonetheless more realistic. This conclusion is illustrated in figure~\ref{fig:limits_age_comparison}, where we overlay the stacking bounds by assuming equilibrium and real age:  The difference amounts to less than 30\%.

\section{Discussion and conclusions}
\label{sec:conclusions}

\begin{figure}
\centering
 \includegraphics[width=0.85\linewidth]{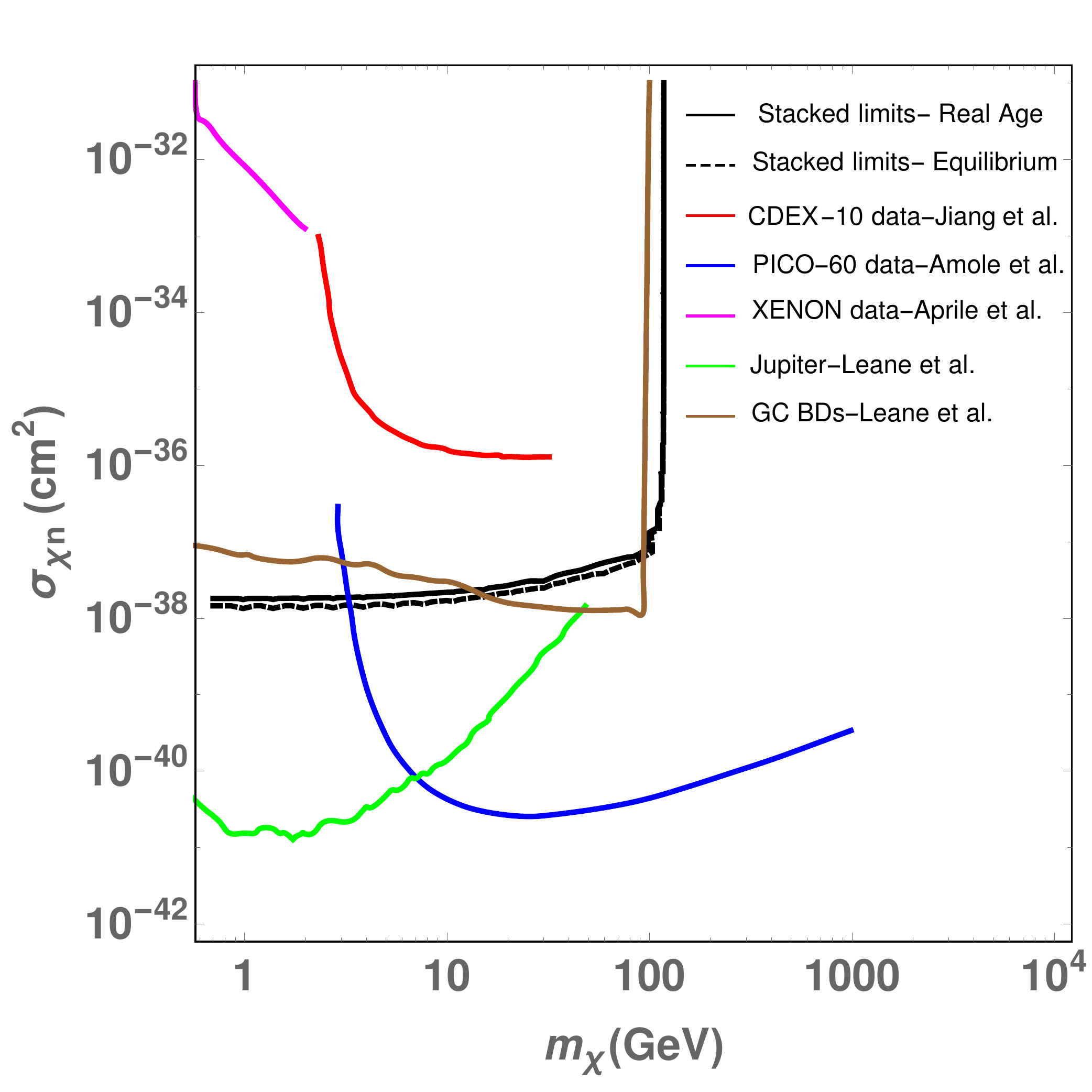}
\caption{Comparison between the limits obtained from our work with other literature studies. We show the constraints from direct detection on the 
DM-proton spin dependent cross section~\cite{CDEX:2018lau, PICO:2019vsc, XENON:2019zpr}, as well as from DM annihilation to long-lived
particles for BDs in the Galactic center \cite{Leane:2021ihh} and from the Jupiter \cite{Leane:2021tjj}.
}
 \label{fig:BD_compare_studies}
\end{figure}

In the present work, we have considered a sample of nearby ($<$ 11 pc distance), cold and old 
BDs, and looked for gamma-ray excess emission from the direction of these objects using
the data from the \Fermi-LAT telescope.

In the absence of any excess, we were able to set 95\% C.L.~upper limits on the 
scattering cross section of DM particles with nucleons as a consequence of
the capture of DM particles within the BD and subsequent annihilation into long-lived mediators.
The latter, while interacting only feebly within the BD, escape the celestial object and then
decay into photons leading to a characteristic box-like energy spectrum.
Besides individual object limits, we also provided the combined bound coming from the stacking of all nine objects in our sample, improving the single most stringent bound by almost one order of magnitude. 

A number of observations are in order: 
\begin{itemize}
    \item In our DM modelling, we have neglected evaporation effects. For this 
    approximation to be correct, the DM mass should be larger than $\sim$ 0.7 GeV 
    for BD systems~\cite{Garani:2021feo}.
    The same reasoning applied to Jupiter limits the range of applicability 
    of the limits to $m_\chi \gtrsim$ GeV, meaning that an extrapolation to 100 MeV mass is 
    incorrect~\cite{Garani:2021feo}.
    We expect BDs and super-Jupiter planets to be able to probe the smallest DM masses within
    this approximation. We, therefore, cut all plots at 0.7 GeV.
    \item We explicitly tested the correctness of the equilibrium hypothesis. Knowing the 
    age of the selected BD, and using models for the inner temperature and density of these objects
    we derived the time required to bring capture and annihilation in equilibrium. 
    We demonstrated that, for the scattering and annihilation cross section of relevance here, the 
    equilibrium hypothesis is well justified. The difference in the limits on the scattering cross section between the case where equilibrium is assumed or not is about a factor of 1.26
    for 1 GeV DM mass. Even if of negligible impact on the sample under study, we recommend 
    using the real age whenever known BDs are used to set such a type of constraint since this
    results in more realistic limits. Similar considerations apply to Jupiter as well.
    \item In case the real age is used, the limits on the DM scattering cross section depend on 
    the value of the annihilation cross section adopted. We here set $\langle \sigma v \rangle$ to
    the current upper limits in this DM mass range set by cosmological observation on velocity 
    independent (s-wave) annihilation processes. We notice that viable DM particle models with
    long-lived mediators typically do predict velocity-dependent annihilation cross sections. 
    In the case of, e.g.~p-wave annihilation, the CMB bounds are not so strong (because of the
    important velocity suppression in the early universe), and more important bounds
    are set with searches for DM in hard X-rays and cosmic rays~\cite{Essig:2013goa,Boudaud:2018oya}.
    Considering the p-wave annihilation cross-section upper limits from~\cite{Boudaud:2018oya} and a
    relative DM-DM velocity within the BD of $\mathcal{O}$(10) km/s, we obtain
    $\mathcal{O}$(10$^3$) lower annihilation cross-section values within the BD, which imply
    an increased equilibration time and so a larger difference between real age and equilibrium limits.
    This is a general observation which also applies to any other limit from 
    celestial objects in the literature. 
    \item Having limited ourselves to a point-like source gamma-ray analysis, we restrict the 
    parameter space for the decay length of the long-lived mediator to a range which is bounded 
    from below by the condition that the mediator should decay outside the BD ($L \gtrsim R_\star$)
    and from above by the point-like source approximation of the signal ($L \lesssim d_\star \theta_{68\%}$). Note that analogous conditions applied to Jupiter, as in~\cite{Leane:2021tjj}, lead to a similar lower limit but to five orders of magnitude more stringent upper limit, controlled by the significant hierarchy in distance between $\sim 5$AU (typical distance to Jupiter) and a few pc (distance to the nearest BDs). 
    Hence, even when our bounds are not naively competitive to those reported in~\cite{Leane:2021tjj}, they apply to much wider parameter space. 
    We also comment here about limits from the Sun, set through the same DM signal modelling~\cite{Mazziotta:2020foa, HAWC:2018szf, Leane:2017vag}: 
    While these limits are about 8 o.d.m.~stronger than ours, they are valid only for masses above 10 
    GeV and for $L \simeq R_\odot$, i.e.~a much narrower parameter space. We also notice here that our range of applicability is set  by the decay length, $L$. This is linked to fundamental parameters, such as the coupling with photons $g_{\phi\gamma \gamma}$, via the lifetime $\tau$, since $L = (m_\chi/m_\phi) \, c \, \tau(m_\phi, g_{\phi\gamma \gamma})$. Independent constraints in the plane $(m_\phi, g_{\phi\gamma \gamma})$ exist for such mediators, notably from astrophysical and cosmological considerations on axion-like particles, see e.g. the compilation in~\cite{AxionLimits}.
    While we tested that some points can easily evade such constraints, it is clear for instance that not all values of the range $L$ are allowed at each DM mass or that at the lightest DM masses probed viable scenarios may require breaking the approximation $m_\phi\ll m_\chi$, which is unnecessary, but done here for simplicity. A systematic scan of the multi-dimensional parameter space would be needed for more quantitative considerations. 
    
    \item The bounds derived from {\it known} nearby BDs do not
    suffer from the large astrophysical uncertainty on BD distribution or the DM spatial density profiles. Indeed, 
    depending only on the local DM matter density, our limits are only affected by an overall 
    $\sim$ 30\% uncertainty. On the other hand, DM constraints obtained in~\cite{Leane:2021ihh} rely both on the assumed DM density profile and the model of the BD population towards the Galactic center region. 
    We checked that we can reproduce the results of~\cite{Leane:2021ihh} for the adopted Navarro-Frenk-White profile, but also that adopting a Burkert profile with parameters from~\cite{Calore:2022stf}, the 
    GC limits are degraded by about an order of magnitude.
\end{itemize}

Besides astrophysical bounds, we can also compare our results with DM direct detection experiments. These experiments are much more sensitive to spin-independent cross sections than to spin-dependent ones, because the former cross sections benefit of the coherent enhancement on large nuclei used as typical targets. On the contrary, the bounds from BDs are almost equally strong for spin-dependent or spin-independent cross sections, since BDs are mostly made of hydrogen. 
It is therefore more interesting to compare our bounds with the direct detection limits  
obtained by CEDX~\cite{CDEX:2018lau}, PICO-60~\cite{PICO:2019vsc}  and 
XENON1T~\cite{XENON:2019zpr} on the DM-proton spin-dependent scattering cross section. 
Compared with current direct detection bounds, BDs limits have the unique advantage to extend to 
masses lower than a few GeV with sensitivity reaching cross section values of at least $10^{-38}$ cm$^{2}$.

Let us conclude with some perspectives for future improvements:
From the particle physics side, it would be straightforward to generalise the analysis to different final states/spectrums, or to a concrete model of DM coupled to the SM via a mediator. If motivated, it may even be possible to extend the analysis to a longer decay range (by one or two orders of magnitudes) which would require a dedicated extended source analysis of \Fermi-LAT data.

Another avenue to get stronger, yet robust, constraints on sub-GeV particle DM models via BDs would be to look at the cumulative emission from the local BD population. As an example, ref.~\cite{Reyle:2010xw} derived the space density of ultracool-field BDs, for the T and Y type dwarfs, to be around $8.3^{+9.0}_{5.1}$ $\times$ $10^{-3}$ objects $\rm pc^{-3}$, which may imply an interesting sensitivity. 
Also, this high space density indicates that numerous late-type BDs are yet to be detected. In the near future, with deeper and more sensitive optical and infrared sky surveys, such as e.g. via JWST, the detection of ultracool BDs would be increased and that would positively improve the sensitivity of our analysis.

Finally, it remains to be seen if alternative mechanisms for GeV emission of BDs, either exotic or astrophysical ones, can be envisaged
and may motivate future gamma-ray dedicated analyses, as done, e.g., for young red dwarfs~\cite{Song:2020ckf}. Definitely, we should not discard the possibility that 
\Fermi-LAT can still reveal some surprises!

\begin{acknowledgments}
\end{acknowledgments}
The work of P.B. has been supported by the EOSC Future project which is co-funded by the European Union Horizon Programme call INFRAEOSC-03-2020, Grant Agreement 101017536.
\bibliography{bibliography,biblio_evapmass}

\newpage

\onecolumngrid

\section*{Erratum}

\noindent In this Erratum, we report on a mistake which affects the conclusions of the second part of our paper, from Sec. IV onwards, i.e., its model-dependent application to dark matter (DM) phenomenology. We emphasize that the model-independent \Fermi-LAT analysis of gamma rays from brown dwarfs (BDs) as well as the inferred flux bounds, reported in Sec. III, are unaffected.

\noindent In particular, following an update of the fundamental constants block of the code, an error was introduced in the unit conversion factor of the Newton constant G between the natural and the International System of Units (SI units), artificially enhancing the calculation of the capture rate by up to 2 orders of magnitude.  A secondary, technical consequence is that an approximation for the calculation of the probability function $p_{N}$ (i.e. $p_N\simeq p_{N1}$= $\sigma_{\chi n} / \sigma_{\rm sat}$, holding for $\tau \ll 1$) valid for the previously implemented erroneous numerical value, is now unsuitable, and users of the numerical scripts must resort to the complete formula for $p_{N}$.

\noindent Accounting for this correction, current \Fermi-LAT upper limits on the gamma-ray flux of the selected BDs do not enable to set a bound on the DM-nucleon cross-section. An improvement of approximately a factor 9 in the gamma-ray limits is required in order to obtain bounds on the scattering cross-section $\sigma_{\chi n}$ $\sim 10^{-36}$ cm$^{2}$ for DM masses below 10 GeV, and an improvement of roughly two orders of magnitude are needed to lead to bounds comparable to those shown in our original publication. 

\noindent We thank Thong Nguyen (Stockholm University) for spotting some inconsistency in the publicly available code that triggered this Erratum. 



\end{document}